\documentclass{emulateapj}
\submitted{{\it Accepted for publication in AJ}}
\usepackage{epsfig}
\shortauthors{West et al.}
\shorttitle{Activity and Dynamics of M7 Dwarfs}
\begin{document}

\title{Using the Galactic Dynamics of M7 Dwarfs to Infer the Evolution of Their Magnetic Activity}

\author{Andrew A. West\altaffilmark{1,2,3}, 
John J. Bochanski\altaffilmark{3},
Suzanne L. Hawley\altaffilmark{3}, 
Kelle L. Cruz\altaffilmark{4},
Kevin R. Covey\altaffilmark{3},
Nicole M. Silvestri\altaffilmark{3},
I. Neill Reid\altaffilmark{5},
James Liebert\altaffilmark{6}}

\altaffiltext{1}{Corresponding author: awest@astro.berkeley.edu}
\altaffiltext{2}{Astronomy Department, University of California, 601 Campbell Hall, Berkeley, CA 94720-3411}
\altaffiltext{3}{Department of Astronomy, University of Washington, Box 351580,
Seattle, WA 98195}
\altaffiltext{4}{American Museum of Natural History, Department of Astrophysics, Central Park West at 79th St., New York, NY 10024}
\altaffiltext{5}{Space Telescope Science Institute, 3700 San Martin Drive,Baltimore, MD 21218}
\altaffiltext{6}{Department of Astronomy and Steward Observatory, University of Arizona, Tucson, AZ 85721}

\begin{abstract} We present a spectroscopic study and dynamical analysis of $\sim$2600 M7 dwarfs.  We confirm our previous finding that the fraction of magnetically active stars decreases with vertical distance from the Galactic plane.  We also show that the mean luminosity of the H$\alpha$ emission has a small but statistically significant decrease with distance.  Using space motions for $\sim$1300 stars and a simple one-dimensional dynamical simulation, we demonstrate that the drop in the activity fraction of M7 dwarfs can be explained by thin disk dynamical heating and a rapid decrease of magnetic activity at a mean stellar age of $\sim$6-7 Gyr. 

\end{abstract}


\section{Introduction}

The most ubiquitous stars in the Galaxy are low-mass dwarfs.  Because
they are plentiful, they provide an excellent probe of the structure
and dynamics of the local thin disk stellar population.  In addition,
many low-mass dwarfs are magnetically active (as identified by
H$\alpha$ emission), especially the mid to late M-dwarfs.  Various
studies have suggested that there is a relation between the magnetic
activity of an M-dwarf and its age.  These studies have used the
Galactic motions of M-dwarfs (Wilson \& Woolley 1970; Wielen 1974;
Hawley et al. 1996), M-dwarf activity in stellar clusters (Hawley et
al. 1999; Hawley et al. 2000; Gizis, Reid \& Hawley 2002) and active
M-dwarfs in binary systems (Silvestri, Hawley \& Oswalt 2005;
Silvestri et al. 2006). By utilizing the activity of M-dwarfs, and the
largest spectroscopic sample of M-dwarfs ever assembled, we can
augment dynamical and positional data with the additional parameter of
time.

Previous studies have thoroughly investigated the vertical dynamics
and structure of the Galaxy. Wielen (1977) found that the velocity
dispersion of stars increases with time to the 0.5 power (t$^{0.5}$).
Spitzer \& Schwarzchild (1951, 1953) suggested that massive gas clouds
would cause an increase in the velocity dispersion of stars through
cloud-star interactions.  Subsequent observations of molecular clouds
confirm that enough interstellar gas exists to dynamically heat disk
stars.  However, current evidence indicates that molecular clouds
alone are not responsible for the total amount of heating.  Black
holes and spiral density waves have been suggested as other possible
contributors to the heating of the Galactic disk (Jenkins 1992;
H\"anninen \& Flynn 2002).

Recently, many groups have investigated the structure and age of the
Galactic Disk by examining the observed and simulated kinematics of nearby
stars (Kuijken \& Gilmore 1989; Dehnen \& Binney 1998; Rocha-Pinto \&
Maciel 1998; Binney, Dehnen \& Bertelli 2000; Fuchs et al. 2001;
H\"anninen \& Flynn 2002; Holmberg et al. 2003; Siebert, Bienayme \&
Soubiran 2003).  The values reported for the exponent ($\alpha$) of
the time dependence (t$^{\alpha}$) of the velocity dispersion have not
converged and range from 0.26-0.59 (in most cases the uncertainties in
$\alpha$ are smaller than this spread).  Despite the discrepancies,
these studies have aided in constraining the age of the solar
neighborhood, the surface mass density of the Galactic plane and the
time evolution of dynamical heating.  Most of this work has not
utilized low-mass stars due to the small numbers of M-dwarfs included
in magnitude-limited samples.  Aside from the initial work by Wielen
(1974, 1977) and the more recent PMSU and 100 Parsec surveys (Reid,
Gizis \& Hawley 2002; Bochanski et al. 2005), few Galactic studies
have examined the low-mass stellar constituents.

The advent of large survey science has vastly increased the number of
low-mass dwarfs available for analysis.  Using several hundred M7
dwarfs from the Sloan Digital Sky Survey (SDSS), West et al. (2004;
hereafter W04) suggested that the fraction of activity in M7 stars
decreases as a function of height above (or below) the Plane.  They
suggested that this phenomenon would naturally result from an
age-activity relation; the older stars are on average farther away and
are no longer active, thus decreasing the active fraction.  However,
despite using hundreds of stars in their analysis, the W04 results had
large uncertainties.

In this paper, we focus on the activity and dynamics of a much larger
sample of M7 dwarfs.  We describe our selection criteria and data in
\S2.  In \S3 we confirm the findings of W04, namely that the fraction
of activity decreases as a function of vertical distance from the
Plane. We also investigate how the magnitude of activity (as
quantified using L$_{\rm{H\alpha}}$/L$_{bol}$) changes with distance.
We utilize proper motions from the USNO-B catalog (Monet et al. 2003)
and a new set of low-mass star templates for measuring radial
velocities (Bochanski et al. 2006) to derive space motions. In \S4
we use a simple dynamical simulation to explain the decline in
activity fraction as due to dynamical heating effects, and derive a
timescale for the existence of magnetic activity in M7 dwarfs.  Our
results are discussed and plans for future projects proposed in \S5.

\section{Data}

The SDSS (Gunn et al. 1998; Fukugita et al. 1996; York et al. 2000;
Hogg et al. 2001; Gunn et al. 2006; Ivezi{\'c} et al. 2004; Pier et
al. 2003; Smith et al. 2002; Stoughton et al. 2002) provides ideal
samples to examine the spectral and photometric properties of low-mass
dwarfs.  Many studies have already taken advantage of the
unprecedented large sample sizes and uniform data quality that SDSS
offers for cool-star research (Hawley et al. 2002; W04; Silvestri et
al. 2006).  In this study, we continue the analysis originally
discussed in W04, concentrating on M7 dwarfs.  This spectral type was
selected because of the large number of M7 dwarfs in the sample and
their high activity fraction.

Our sample is drawn from the SDSS Data Release 4 (DR4;
Adelman-McCarthy 2006). A total of 4420 stars were selected from the
SDSS spectroscopic database based on the colors typical of M7 dwarfs
(1.92 $< r-i <$ 2.65 and 1.06 $< i-z <$ 1.45; West, Walkowicz, \&
Hawley 2005). Because there is some overlap with M6 and M8 stars, this
sample contains all three spectral types but is dominated by M7s.
Nearby M7 dwarfs are too bright for SDSS spectroscopy.  Therefore, 171
additional M7 dwarfs were added from the nearby 2MASS selected sample
of Cruz et al. (2003) to add sensitivity to our sample at low Galactic
heights.  We examined each spectrum by eye using the Hammer stellar
spectral-typing facility (Covey et al. in prep).  We removed 232
spectra because they were identified as galaxies, binary systems or
had poor data quality.

We performed the spectral activity analysis described in W04 on the
entire M7 sample.  There were 2601 stars that had the requisite data
quality (using the same criteria as W04) for a statistically robust
analysis. The resulting sample contains 3 times more M7 dwarfs than
that of W04.  We found that 1828 of the stars were active and 773 were
not active, giving a 70$\pm$2\% active fraction, similar to the
64$\pm$3\% M7 activity fraction found in W04.  Some of the discrepancy
is due to the addition of the Cruz et al. (2003) sample that consists
almost entirely of active stars (because they are nearby).  Using the
method of Walkowicz, Hawley \& West (2004), we also computed the ratio
of luminosity in the H$\alpha$ line to the bolometric luminosity
(L$_{\rm{H\alpha}}$/L$_{bol}$) for all stars in the sample.  We
averaged the $\chi$ values for the M6.5, M7, M7.5 and M8 spectral
types for determining the L$_{\rm{H\alpha}}$/L$_{bol}$
ratios. Distances to all stars were calculated using the photometric
parallax methods of Walkowicz et al. (2004) and West et al. (2005).
Using the positions and distances of each star, Galactic height was
computed assuming the Sun is 27 pc above the Plane (Chen et al. 2001).

\begin{figure}
\includegraphics[angle=90,scale=.38]{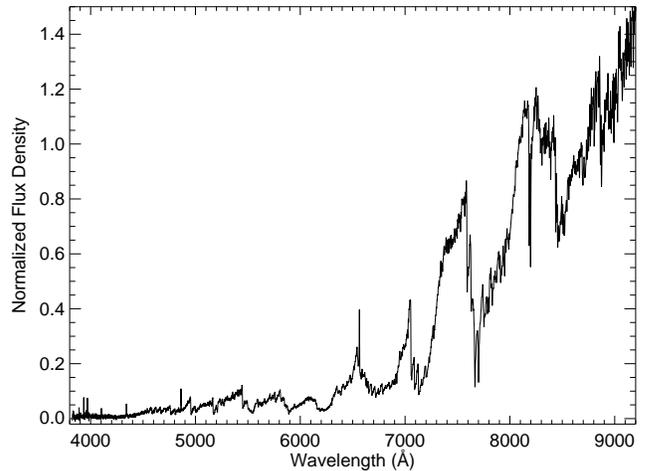}
\caption{Co-added template spectra for all SDSS stars in our M7 sample.  Spectra are shifted to zero velocity and placed on a finer resolution grid to produce more accurate radial velocities (see Bochanski et al. (2006) for more details).} 
\label{fig:template} 
\end{figure}

We utilized the SDSS/USNO-B matched catalog to obtain proper
motions. The USNO-B proper motions were obtained for 1180 of the 2601
good quality SDSS stars.  To maintain a uniform sample of space
motions, we did not include the Cruz et al. (2003) data in our
dynamical analysis.  Radial velocities were measured for all SDSS
stars using the method described in Bochanski et al. (2006), by which
co-added, zero-velocity SDSS spectra were used to create high
signal-to-noise ratio (SNR) radial velocity templates. Figure
\ref{fig:template} shows the resulting template from the co-addition
of all stars in our M7 sample.  Our derived radial velocities are
accurate to within $\sim$5 km s$^{-1}$.  Using these proper motions and
radial velocities, space motions were computed for all SDSS stars.

\begin{figure}[h]
\includegraphics[scale=.49]{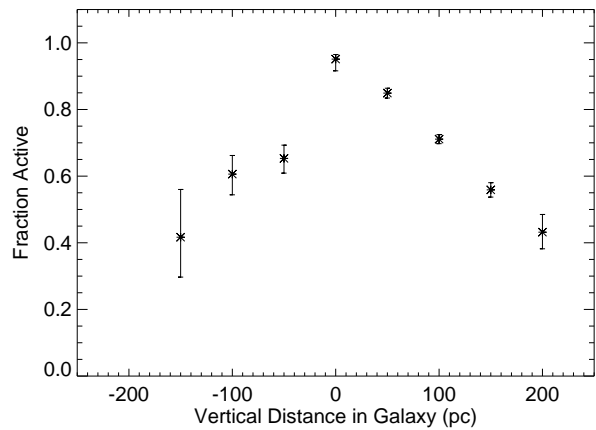}
\caption{The fraction of active stars is shown as a function of vertical distance above and below the Galactic Plane.  The stars have been binned every 50 pc.  Active stars are concentrated toward the Plane and the fractions decreases on either side. Using an age-activity relation and assuming dynamical heating, we infer that the stars closer to the Plane are younger and that the decline in activity fraction is likely due to the termination of magnetic activity in older stars.} 
\label{fig:m7frac} 
\end{figure}

\section{Observational Results}

\begin{figure}
\includegraphics[scale=.49]{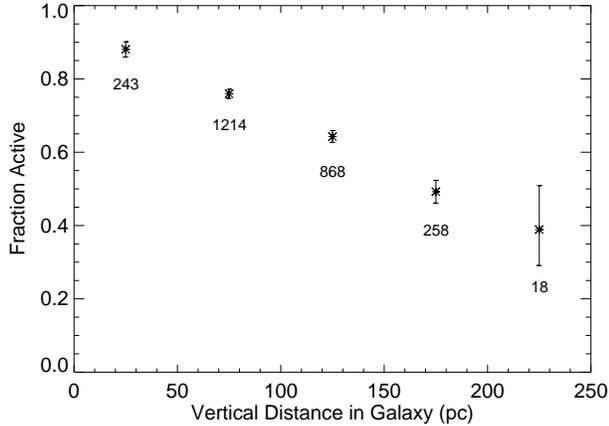}
\caption{Same as Figure \ref{fig:m7frac} with the stars folded across the Plane in order to obtain a higher signal-to-noise result.  The values below each point represent the number of stars in each bin.} 
\label{fig:m7fracwrap} 
\end{figure}

Figure \ref{fig:m7frac} shows the fraction of active M7 dwarfs, in 50
pc bins as a function of height from the Galactic plane. This figure
confirms the result of W04; the fraction of active M7s decreases as a
function of distance from the Plane.  Because these data are binomial
in nature, the uncertainties are asymmetric and were calculated using
the binomial-distribution method described in Burgasser (2003).  For
bins with large numbers of stars ($\gtrsim200$),the uncertainties are
symmetric and well described by Poisson statistics\footnote{Note that
the final equation in the Burgasser (2003) appendix describing the
Poisson uncertainty has a typographical error; the ``$+$'' symbol
should be replaced with a ``$-$'' and should read
($\epsilon_{b}^{U}-\epsilon_b)/\epsilon_b=(\epsilon_{b}^{L}-\epsilon_b)/\epsilon_b=(1/n-1/N)^{1/2}$}. These
uncertainties are substantially smaller than those reported by W04 due
to the larger sample. Figure \ref{fig:m7fracwrap} shows the same data
folded across the Galactic Plane in order to increase the SNR. The
numbers below each data point indicate the number of stars in each
bin.

Figures \ref{fig:m7frac} and \ref{fig:m7fracwrap} confirm that the
activity fraction decrease is real.  In the binned data, the stars
that are further away from the Plane have experienced more dynamical
heating and are thus older.  As discussed above, the decrease in
activity fraction may thus be due to an age-activity relationship.  In
\S4 we quantify this suggestion using a simple dynamical simulation.

\begin{figure}
\includegraphics[scale=.49]{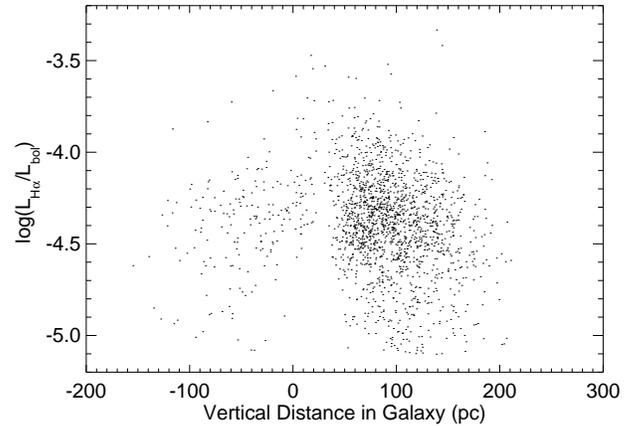}
\caption{L$_{\rm{H\alpha}}$/L$_{bol}$ for all stars in the sample as a function of distance above and below the Galactic Plane.  Because we investigate only a single spectral type in this study, the L$_{bol}$ value is nearly the same for every star.  Therefore, this figure demonstrates the large spread in H$\alpha$ luminosity at all distances.} 
\label{fig:lhalbol} 
\end{figure}

\begin{figure}[h]
\includegraphics[scale=.49]{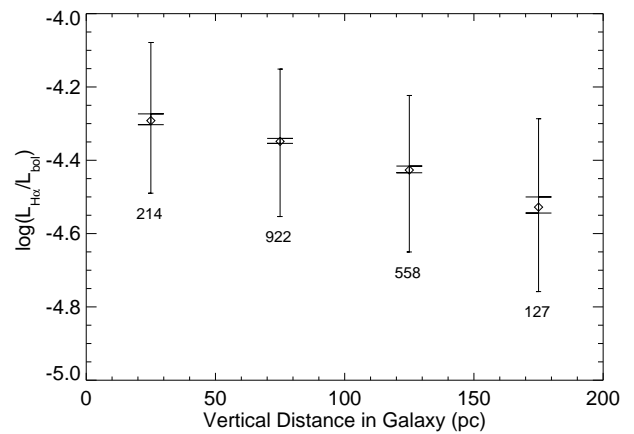}
\caption{ Mean L$_{\rm{H\alpha}}$/L$_{bol}$ as a function of absolute height above the Plane (in 50 pc bins).  The numbers below each point indicate the number of active stars in each bin.  The narrow error bars represent the 1$\sigma$ spread of the bin, while the wide error bars indicate the uncertainty in the mean.  The mean does decrease as a function of vertical height, indicating a small, but significant decrease in activity with age. In addition, the large spread at all distances and the lack of small L$_{\rm{H\alpha}}$/L$_{bol}$ values at close distances argues that activity may suddenly cease when an M-dwarf reaches a given age.} 
\label{fig:lhalbolmean} 
\end{figure}

We also investigated how the L$_{\rm{H\alpha}}$/L$_{bol}$ ratio
changes as a function of vertical height above the Plane (Figure
\ref{fig:lhalbol}) for the active stars in the sample. The
L$_{\rm{H\alpha}}$/L$_{bol}$ values have a very large scatter.
Because the stars in this sample have nearly the same bolometric
luminosity, this scatter represents the scatter in H$\alpha$
luminosity for M7 dwarfs.  In order to investigate any trend with
distance, we again binned the data and folded them across the Galactic
plane.  Figure \ref{fig:lhalbolmean} shows the resulting relation.
The narrow error bars indicate the rms scatter of the data in each
bin, while the wide error bars represent the uncertainty in the mean.
There is a small but highly significant decrease in the mean
L$_{\rm{H\alpha}}$/L$_{bol}$ ratio as a function of vertical distance,
which clearly demonstrates that magnetic activity decreases with
stellar age. A linear fit to the data revealed a slope of
$-$0.0015$\pm0.00017$ dex $\rm{pc^{-1}}$, indicating this result
has a 10 sigma significance.  Does magnetic activity decline
continuously or does it eventually go through a rapid decline?  The
lack of stars with low L$_{\rm{H\alpha}}$/L$_{bol}$ ratios at close
distances and the presence of high SNR inactive stars spread over all
Galactic heights suggests that M7 dwarfs might eventually shut off
rapidly rather than undergoing a perpetually slow decline.

With the space motions derived from the radial velocities, proper
motions and distances, we examined how the vertical velocity
dispersion changes as a function of height (Figure \ref{fig:simsig}).
The vertical velocity dispersion has been determined in previous
studies (e.g. Wielen 1977; Siebert et al. 2003; Bochanski et
al. 2005). However, our data represent the largest sample of low-mass
dwarfs yet used for Milky Way kinematic analysis, and therefore yield
the most statistically significant result.  We confirm that there is a
steady rise in the velocity dispersion as stars are dispersed away
from the Galactic plane.

\section{Simulations}

We proceeded to test the hypothesis that the magnetic activity
fraction decrease is caused by a rapid decline in activity in the
older, more dynamically heated stars.  We developed a simple
one-dimensional model to trace the vertical dynamics of stars as a
function of time, using the ``leap-frog'' integration technique (Press
et al. 1992) and the vertical Galactic potential from Kuijken \&
Gilmore (1989) and Siebert et al. (2003) given by:

\begin{equation}
\Phi(z)=2\pi G\left(\Sigma_{0}\sqrt{z^2+D^2}-D)+\rho_{\rm eff}z^2\right),
\end{equation}

\noindent where $\Sigma_{0}$ is total surface mass density, D is the mass scale height and $\rho_{eff}$ is the halo local effective mass density.  We use the values given in the middle column of Table 3 in the Siebert et al. (2003) study for these quantities.

We assumed a constant star formation rate, and injected a new
population of 50 stars at the Galactic midplane, every 200 Myr, for a
total simulation time of 10 Gyrs.  Each new group of stars began with
a randomly drawn velocity dispersion of 8 km s$^{-1}$ (Binney et
al. 2000) and had a new position and velocity computed every 0.1 Myr.
We simulated dynamical heating by altering the velocities (energies)
of stars such that their new velocity dispersions would keep them in
agreement with the t$^{0.5}$ relation described above (Wielen 1977;
Fuchs et al. 2001; H\"anninen \& Flynn 2002).  Energy was added only
when stars were within a given distance from the Plane.  These
distances were symmetric about the midplane and were varied from
$\pm$3 pc to $\pm$30 pc in intervals of 3 pc.  Our simulation tracked
the velocity, position and age of each star in the simulation.

In the Milky Way, molecular clouds and other objects that cause
dynamical heating do not lie in a continuous region that is symmetric
about the Plane.  However, the distances within which energy is added
in our simulations can be considered ``regions of influence,'' and
represent the total distance over which a star is subject to dynamical
heating during its orbit.  Therefore, our simulation reproduces the
effect of many finite interactions spread throughout a star's orbit in
a manner that is computationally simple.

After the dynamical simulations are complete, we introduce a timescale for magnetic activity to the simulated data.  We assign a range of mean activity lifetimes to every star in all simulations and allow these lifetimes to vary from 5 Gyr to 9 Gyr in 0.5 Gyr intervals. The activity lifetimes are drawn from Gaussian distributions with 1$\sigma$ spreads of 1 Gyr. From the resulting data, we can derive activity fractions as a function of vertical distance from the Plane.

\begin{figure}
\includegraphics[scale=.55]{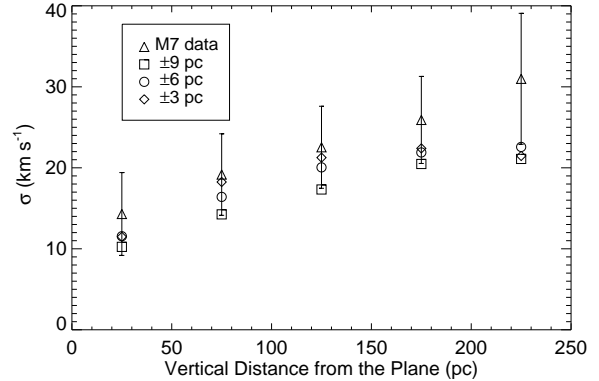}
\caption{Vertical velocity dispersion as a function of distance from
the Plane from our simulation (diamonds, squares and circles) and the
M7 data (triangles).  The three different symbols in the figure are
the three best matched ``regions of influence'' (where dynamical
heating is important) of $\pm$9 pc (squares), $\pm$6 pc (circles) and
$\pm$3 pc (diamonds) in our heating simulation. All of the regions
have been modeled to be symmetric about the Plane for simplicity. The
resulting dispersions agree with the observational data to within the
uncertainties.}
\label{fig:simsig} 
\end{figure}

Figure \ref{fig:simsig} shows the vertical velocity dispersion as a
function of distance from the Plane from our simulation (diamonds,
squares and circles) and the M7 data (triangles).  The three different
symbols in the figure are the three best matched ``regions of
influence'' in our heating simulation.  The observations and
simulations agree to within the uncertainties.  Figure
\ref{fig:simsig} is included to demonstrate that our simple integrator
does indeed produce dynamical outputs that agree with observations.

\begin{figure*}
\epsscale{1}
\plotone{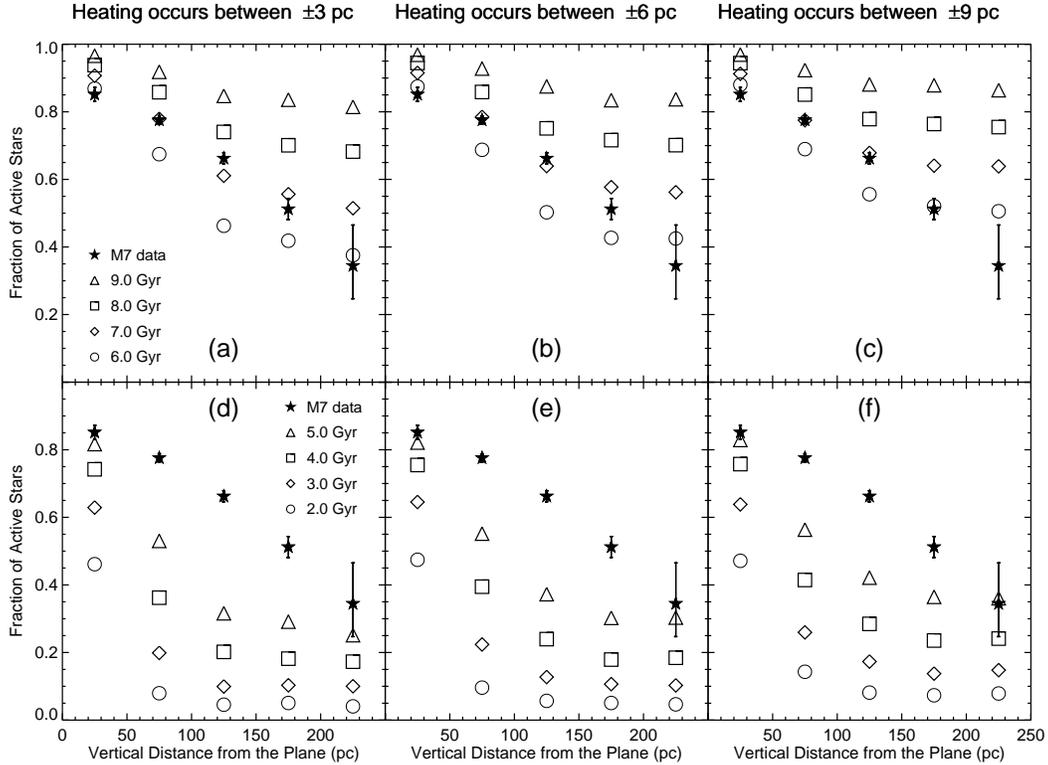}
\caption{Fraction of active stars as a function of absolute vertical distance
from the Plane from our simulations and the M7 sample.  Each column
contains data from a simulation with one of the three regions of
influence shown in Figure \ref{fig:simsig}; $\pm$3 pc (a,d), $\pm$6 pc
(b,e) and $\pm$9 pc (c,f).  The activity fractions as a function of
absolute vertical distance from the Plane are plotted for activity
lifetimes of 6 ,7, 8 and 9 Gyr (top panels) and 2, 3, 4 and 5 Gyr
(bottom panels). The M7 data (stars) are plotted in all 6 panels.  All
simulations show a clear decrease in activity as a function of
distance from the Plane.  The simulations with activity lifetimes
between 6 and 7 Gyr and a $\pm$3 pc ``region of influence'' best match
to the data.} 
\label{fig:simplot} 
\end{figure*}

The dynamical simulations confirm the plausibility of our hypothesis;
dynamical heating and a rapid decrease in magnetic activity at a given
age can explain the activity fraction decrease, as shown in Figure
\ref{fig:simplot}.  Each column contains data from a simulation with
one of the three regions of influence shown in Figure
\ref{fig:simsig}, $\pm$3 pc (a,d), $\pm$6 pc (b,e) and $\pm$9 pc
(c,f).  The activity fractions as a function of absolute vertical
distance from the Plane are plotted for activity lifetimes of 6 ,7, 8
and 9 Gyr (top panels) and 2, 3, 4 and 5 Gyr (bottom panels). The M7
data (stars) are plotted in all 6 panels.  All simulations show a
clear decrease in activity as a function of distance from the Plane.
The simulations with activity lifetimes between 6 and 7 Gyr agree
remarkably well with the observational data considering our very
simple model.  The $\pm$3 pc ``region of influence'' is the best match
to the data.

These results suggest that the presence of magnetic activity in an M7
dwarf depends on the age of the star.  After roughly 6-7 Gyr, M7
dwarfs experience a rapid decrease in their activity such that
activity becomes undetectable.  Throughout their lifetime, M7 dwarfs
are dynamically heated as they move through the disk.  The integral
over the region(s) of space in which heating occurs is only a few
parsecs but is sufficient to match the observed kinematics.

\section{Discussion}

Using over 2600 M7 dwarf spectra, we confirm the result of W04, that
the activity fraction does decrease as a function of vertical distance
from the Galactic Plane.  A possible explanation for this observation
is that activity ceases in older stars and that due to dynamical
heating, these older stars are on average further away.

Our model simulation using simple dynamical arguments and the
assumption that activity ceases at a given stellar age adequately
reproduces the observations and validates the feasibility of our
hypothesis.  A more detailed dynamical simulation should be carried out to aid in the
physical interpretation of the observations. The importance of our
simulation is simply to demonstrate that our hypothesis is reasonable.

The extrapolation of the simple log-linear fit of the age at which activity ceases vs. spectral type (color, mass) derived from a few open clusters in 
Hawley et al (2000) would predict a very large age for the M7 
dwarfs to cease their activity.  Our sample has allowed us to actually 
probe these late-type dwarfs, which are not accessible in old clusters, 
and provides the first measured estimate of the time dependence of their 
magnetic activity.  The investigation of the entire M dwarf sequence by 
our method will allow us to further refine and extend the relation from 
the open clusters (Hawley et al 2006 in prep.). 

The L$_{\rm{H\alpha}}$/L$_{bol}$ data show a large scatter that may be
due to: 1) an intrinsic variation in the activity of a sample of
stars; or 2) significant variation in the H$\alpha$ line emission
during the lifetime of an individual M7 dwarf.  A large sample of
low-mass stars with large time baselines is needed to distinguish
between these possibilities.  If individual stars cannot produce the
variation we see in large single epoch data, then the scatter is
likely the intrinsic scatter of the population.

The L$_{\rm{H\alpha}}$/L$_{bol}$ data suggest that there is a small
but robustly detected decrease in the amplitude of magnetic activity
as a star ages but the large spread at all distances and small
decrease of the mean furthers the notion that these stars have some
range of magnetic activity during their ``active'' lifetimes. The
mechanism behind this age-activity relation is currently
unknown. Future work will statistically explore the spread in
L$_{\rm{H\alpha}}$/L$_{bol}$ to determine if the data support the
rapid decrease of activity hypothesis.


Our results are important for placing constraints on models of
magnetic dynamos in low-mass dwarfs.  Many recent studies have
investigated the mechanisms controlling magnetic field generation in
low-mass stars (e.g. Dobler, Sitx \& Brandenburg 2006; Chabrier \&
K\"uker 2006; Bercik et al. 2006; Donati et al. 2006).  Future dynamo
models should include an age dependence in order to match our
observations.

Work is underway to expand our M7 analysis to include \emph{all} M
dwarf spectral types.  Hawley et al. (in prep) are using the largest
spectroscopic sample of low-mass stars ever assembled (from SDSS Data
Release 5) to examine how the activity fractions, space motions and
L$_{\rm{H\alpha}}$/L$_{bol}$ vary as a function of Galactic height
through the M spectral sequence.  This will allow us to examine how
the dependence of age on activity varies with mass, putting even
stronger constraints on dynamo models.
  
One element that we do not address in this paper is how metallicity
varies as a function of Galactic height.  The sample size of known M7
subdwarfs is quite small and does not allow for a robust statistical
study.  However, the Hawley et al. sample will also allow for an
exploration of how M-dwarf abundances vary with vertical height from
the Plane.  A metallicity gradient, together with strong age
constraints, will provide important information on the chemical
evolution of the Galactic disk.

Almost 30 years have passed since Wielen (1977) used low-mass dwarfs
to probe the kinematics and structure of the Milky Way.  The advent of
large surveys exemplified by SDSS and 2MASS, coupled with the large
abundance of low-mass dwarfs, allows us to continue probing the
structure of the Galaxy with its smallest and most numerous
constituents.

\section{Acknowledgments}
AAW acknowledges the support of NSF grant 0540567, the financial
support of Julianne Dalcanton and the Theodore Jacobsen Fund. SLH and
JJB are supported by NSF grant AST02-05875.  KRC and JJB are supported
by NASA ADP grant NAG5-13111.  KRC is supported by a NASA GSRP
fellowship.  KLC is supported by a NSF Astronomy and Astrophysics
Postdoctoral Fellowship under AST04-01418. The authors would like to
thank Matthew Browning, Lucianne Walkowicz, Adam Burgasser, Gibor
Basri, Anil Seth, Michael Blanton, David Hogg, Beth Willman, Anskar
Reiners and Tom Quinn for useful discussions while completing this
project.  We would like to give a special thanks to the referee for his/her comments that were very helpful in improving this paper.

    Funding for the SDSS and SDSS-II has been provided by the Alfred P. Sloan Foundation, the Participating Institutions, the National Science Foundation, the U.S. Department of Energy, the National Aeronautics and Space Administration, the Japanese Monbukagakusho, the Max Planck Society, and the Higher Education Funding Council for England. The SDSS Web Site is http://www.sdss.org/.

    The SDSS is managed by the Astrophysical Research Consortium for the Participating Institutions. The Participating Institutions are the American Museum of Natural History, Astrophysical Institute Potsdam, University of Basel, Cambridge University, Case Western Reserve University, University of Chicago, Drexel University, Fermilab, the Institute for Advanced Study, the Japan Participation Group, Johns Hopkins University, the Joint Institute for Nuclear Astrophysics, the Kavli Institute for Particle Astrophysics and Cosmology, the Korean Scientist Group, the Chinese Academy of Sciences (LAMOST), Los Alamos National Laboratory, the Max-Planck-Institute for Astronomy (MPIA), the Max-Planck-Institute for Astrophysics (MPA), New Mexico State University, Ohio State University, University of Pittsburgh, University of Portsmouth, Princeton University, the United States Naval Observatory, and the University of Washington.




\end{document}